\documentclass{article}
\usepackage{graphicx,epsfig,rotating,amssymb}

\newlength{\figwidth}
\newcommand\GeV{\ifmmode {\mathrm{\ Ge\kern -0.1em V}}\else
                   \textrm{Ge\kern -0.1em V}\fi}
\begin{document}

\setlength{\unitlength}{1mm} \thispagestyle{empty}

%\begin{center}
%{\Large EUROPEAN ORGANIZATION FOR NUCLEAR RESEARCH}
%\end{center}

%\begin{flushright}
%  {\large CERN-EP/xx-xx }\\
%  {\large DSF  15/99} \\
%  {\large February 14, 2001}\\
%\end{flushright}
\vspace{2cm}

\begin{center}
{\Large \textbf{A combined analysis of all data on $\nu$ and
$\bar{\nu}$ induced single-charm production }}
\end{center}

\vspace{2cm}

%\begin{Authlist}

\begin{center}
\textbf{\large G. De Lellis$^{1)}$, A. Marotta$^{1)}$ and P. Migliozzi$^{2)}$}\\
\vspace{0.2cm} {1) Universit\`{a} Federico II and INFN, Napoli, Italy}\\
{2) INFN, Napoli, Italy}\\

\begin{abstract}
Neutrino and anti-neutrino induced single-charm production are
particularly interesting to study the strange-quark parton
distribution function and the threshold effect in the cross-section,
associated with the heavy quark production. Over the past 30 years,
many experiments have carried out these studies with complementary
techniques: calorimetry, bubble chambers and nuclear emulsions.\\ In
this paper, we review these data and combine them statistically to
extract a world averaged single-charm production cross-section for
both neutrino and anti-neutrino.
\end{abstract}
%
%{\it Submitted to Physics Letters B}
%
\end{center}

%\end{Authlist}
\newpage
%
%%%%%%%%%%%%%%%%%%%%%%%%%%%%%%%%%%%%%%%%%%%%%%%%%%%%%%%%%%%%%%%%%%%%%%%%%%%%%%%
% Introduction
%%%%%%%%%%%%%%%%%%%%%%%%%%%%%%%%%%%%%%%%%%%%%%%%%%%%%%%%%%%%%%%%%%%%%%%%%%%%%%%
%
\section{Introduction}

Neutrino and anti-neutrino induced charm-production is interesting
because it can be used to isolate the strange-quark parton
distribution function and to study the transition to heavy quarks. In
particular, the understanding of the threshold behavior associated
with the charm-production is critical for the extraction of
$\sin^2\theta_W$ from neutrino deep-inelastic data. Furthermore, a
better understanding of the inclusive charm-production cross-section
is crucial for the background determination of future experiments
aiming at the study of neutrino oscillations.

This paper is organized as follow: in Section~\ref{dilepto} we briefly
discuss how charm-production can be studied by looking at the
so-called di-lepton events and review all available di-lepton data. In
Section~\ref{emu} the inclusive charm search by using nuclear
emulsions is discussed together with the available data. In
Section~\ref{combi} the di-lepton and emulsion data are combined to
compute a world averaged inclusive charm-production cross-section.

%
%%%%%%%%%%%%%%%%%%%%%%%%%%%%%%%%%%%%%%%%%%%%%%%%%%%%%%%%%%%%%%%%%%%%%%%%%%%%%%%
% Experimental issues for charm production measurements
%%%%%%%%%%%%%%%%%%%%%%%%%%%%%%%%%%%%%%%%%%%%%%%%%%%%%%%%%%%%%%%%%%%%%%%%%%%%%%%
%
%\section{Experimental issues for charm-production measurements}

%
%%%%%%%%%%%%%%%%%%%%%%%%%%%%%%%%%%%%%%%%%%%%%%%%%%%%%%%%%%%%%%%%%%%%%%%%%%%%%%%
% Charm production studies with di-leptons
%%%%%%%%%%%%%%%%%%%%%%%%%%%%%%%%%%%%%%%%%%%%%%%%%%%%%%%%%%%%%%%%%%%%%%%%%%%%%%%
%
\section{Charm-production studies with di-leptons}
\label{dilepto}
\subsection{Experimental issues}
Many experiments have studied neutrino and anti-neutrino
charm-production by looking at the presence of two oppositely charged
leptons in the final state. This technique was firstly used in
1974~\cite{benvenuti} when the neutrino induced charm-production was
discovered.

In the case of neutrino scattering, the underlying process is a
neutrino charged-current (CC) interaction with an $s$ or $d$ quark,
producing a charm-quark that fragments into a charmed hadron. The
charmed hadron may decay semi-leptonically producing opposite sign
di-leptons through the process:

\begin{eqnarray}
\nu _\mu \;+\;{\rm N}\;\longrightarrow \;\mu ^{-}\!\! &+&\!c\;+\;{\rm X} \\
&&\!\!\hookrightarrow s\;+\;l ^{+}\;+\;\nu _l   \nonumber
\end{eqnarray}                                              

Analogously an anti-neutrino can interact with a $\bar{s}$ or
$\bar{d}$ anti-quark, producing a charm anti-quark that fragments into
a charmed hadron, again leading to a final state with two oppositely
charged leptons.

Since 1974, several experiments~(\cite{cdhs} to~\cite{e616}) have used
this technique to study charm-production. They can be separated in two
classes: calorimetric and bubble chamber experiments, exploiting the
muonic and electronic decay of the charmed hadron, respectively.

Calorimetric experiments are characterized by a massive iron target
and a muon spectrometer to identify the muon and measure its
charge. For these experiments pion and kaon decays constitute the main
background. The high density of the target calorimeter minimizes this
background due to the short interaction length of the detector. A
further background reduction is obtained by requiring a minimum
momentum, typically $p_{\mu}>\mbox{5~GeV}$, for the less energetic
muon. The drawback of such a selection is that these experiments are
not able to search for charm-production at relatively low neutrino
energies. For a typical calorimetric experiment it is not possible to
investigate energy regions below 15~GeV, where the slow-rescaling
threshold effect is more important.

The main characteristic of a bubble chamber filled with a mixture of
heavy liquids (Ne-H$_2$, freon-propane) is its high efficiency in
identifying electrons. Therefore, they searched for charm-production
by looking at $\mu^-e^+$ events. In these experiments, the low
threshold, $p_{e^+}>\mbox{0.3~GeV}$, combined with high statistics for
$E_\nu<\mbox{30~GeV}$, gives good sensitivity to the slow-rescaling
threshold behaviour. The main background sources for these searches
are $\pi^0$ Dalitz decays and $\bar{\nu}_e$CC interactions.

\subsection{Di-lepton data sample and selection criteria}
Table~\ref{tab:dile} summarises the available (anti-)neutrino di-lepton
data samples.

The data selection is similar for all calorimetric experiments,
although the kinematical cuts can be slightly different. In the
following we describe qualitatively the data selection flow:

\begin{itemize}
\item the event must occur in coincidence with the beam and fire the
penetration trigger (CC interaction trigger);
\item to ensure both the longitudinal and transverse containment of
the hadronic shower a fiducial cut is applied;
\item both muons have to be well reconstructed, with the closest
approach between them being less than a certain distance (typically
$\mathcal{O}(10~\mbox{cm})$). This cut allows to reject overlays of
two CC events as well as obvious muons from the decay of shower
hadrons.
\end{itemize}

The di-lepton events are categorised as originating from an incident
neutrino or anti-neutrino by assuming that the primary muon (the muon
produced at the leptonic vertex) is the one with the largest
transverse momentum with respect to the beam direction.

For each event, the reconstructed muon parameters at the vertex
($\overrightarrow{p}_i$, $E_i$,
$\overrightarrow{r}_i=\overrightarrow{p}_i/\mid{p_i}\mid$, $i=1,2$)
and the shower energy $E_{had}$, are used to compute the following
kinematical variables:

\begin{itemize}
\item $E^\nu_{vis}=E_1+E_2+E_{had}$, the visible neutrino energy;
\item
$Q^2_{vis}=2E^\nu_{vis}(E_1-\overrightarrow{p}_1\cdot\overrightarrow{i})-m^2_\mu$,
the visible negative four-momentum trasfer squared, where
$\overrightarrow{p}_1$ is the leading muon 3-momentum and
$\overrightarrow{i}$ represents a unit vector parallel to the beam
direction;
\item $\nu_{vis}=E_2+E_{had}$, the visible energy transferred to the hadronic system;
\item $x_{vis}=Q^2_{vis}/2M\nu_{vis}$, the visible Bjorken $x$;
\item $y_{vis}=\nu_{vis}/E^\nu_{vis}$, the visible Bjorken $y$.
\end{itemize}

In addition to the topological cuts listed above, a set of kinematical
cuts is also applied. Both muons are required to have an energy
greater than 4-10~GeV, primarly to reduce the meson decay
background. The visible hadronic energy is required to be more than
5-20~GeV to ensure a good hadronic energy reconstruction.

The bubble chamber experimental search for charm-production relies
upon its high efficiency in identifying electrons. Therefore, in order
to identify $\mu^-e^+$ dilepton candidates, the film is scanned for
all events with an $e^+$ coming from the vertex with a momentum
greater than $300~\mbox{MeV}$. To be identified as an $e^+$, the track
is required to exhibit at least two characteristics of positrons in
heavy liquid. The signatures are:

\begin{itemize}
\item bremsstrahlung with a $\gamma\rightarrow e^+e^-$ conversion;
\item spiralisation;
\item production of a $\delta$ ray with energy comparable to the primary track;
\item annihilation with two $\gamma\rightarrow e^+e^-$ conversions.
\end{itemize}

After being scanned, measured and reconstructed, each event containing
an $e^+$ track is carefully analyzed, a fiducial volume cut applied
and the search for at least one leaving negative ($L^-$) track
performed. The highest momentum $L^-$ track in each event is
interpreted as a muon leaving the chamber. Under this assumption, a
kinematical analysis is applied in order to select genuine $\nu_\mu$CC
interactions.

\begin{table}[tbp]
\begin{center}
{\small
\begin{tabular}{||c|c|c|c||}
\hline
Experiment   & Technique     & $l^-l^+$ sample &  $l^+l^-$ sample \\
\hline
\hline
CDHS~\cite{cdhs}   & Calorimetry   &  $9922$             & $2123$               \\
LBL Coll.~\cite{lbl}    & Bubble Chamber&  $\mathcal{O}(50)$  &   ---                \\
CBNL Coll.~\cite{bnl}    & Bubble Chamber&     $\mathcal{O}(100)$            &   ---                \\
NOMAD~\cite{nomad}        & Calorimetry   &  $2714\pm227$        & $115^{+38}_{-41}$    \\
CCFR~\cite{ccfr}         & Calorimetry   &  $4247\pm90$        & $944\pm34$          \\ 
Gargamelle I~\cite{garga1} & Bubble Chamber&  $62\pm10$          &   ---               \\
Gargamelle II~\cite{garga2}& Bubble Chamber&  $\mathcal{O}(50)$  &   ---                \\
E53A+E53B~\cite{e53}    & Bubble Chamber&  $602\pm208$        &   ---                \\
Foudas~et~al.~\cite{foudas}       & Bubble Chamber&  $1460.4\pm42.1$    & $223.5\pm5.0$        \\ 
E616+E701~\cite{e616}    & Calorimetry   &  $852\pm77$         & $68\pm16$            \\
\hline
\hline
\end{tabular}
}
\end{center}
\caption{Available (anti-)neutrino di-lepton data samples. The samples are background subtracted.}
\label{tab:dile}
\end{table}

\begin{table}[tbp]
\begin{center}
{\small
\begin{tabular}{||c|c|c|c|c|c||}
\hline
CDHS  & & & LBL & & \\
$E_\nu$ & $<E_\nu>$ & $\sigma_{l^-l^+}/\sigma_{CC}$ & $E_\nu$ & $<E_\nu>$ & $\sigma_{l^-l^+}/\sigma_{CC}$  \\
 (GeV) & (GeV) & (\%) & (GeV) & (GeV) & (\%) \\
\hline
30-40 & 35 & $0.19\pm0.08$ & 0-30 & 15 & $0.21^{+0.65}_{-0.11}$ \\
\hline
40-60 & 50 & $0.40\pm0.11$ & 30-60 & 45 & $0.71\pm0.19$ \\
\hline
60-80 & 70 & $0.53\pm0.10$ & 60-100 & 80 & $0.71\pm0.19$ \\
\hline
80-100 & 90 & $0.60\pm0.11$ & 100-170 & 135 & $0.78\pm0.24$ \\
\hline
100-120 & 110 & $0.61\pm0.11$ & 170-280 & 225 & $0.95\pm0.36$ \\
\hline
120-140 & 130 & $0.69\pm0.13$ & --- & --- & --- \\
\hline
140-160 & 150 & $0.68\pm0.11$ & --- & --- & --- \\
\hline
160-180 & 170 & $0.69\pm0.13$ & --- & --- & --- \\
\hline
180-200 & 190 & $0.73\pm0.13$ & --- & --- & --- \\
\hline
$\>$200-240 & 220 & $0.76\pm0.13$ & --- & --- & --- \\
\hline
\hline
CBNL & & & NOMAD & & \\
$E_\nu$ & $<E_\nu>$ & $\sigma_{l^-l^+}/\sigma_{CC}$ & $E_\nu$ & $<E_\nu>$ & $\sigma_{l^-l^+}/\sigma_{CC}$  \\
(GeV)  & (GeV) & (\%) & (GeV) & (GeV) & (\%) \\
\hline
0-15 & 7.5 & $0.19\pm0.07$ & 0-20 & 10 & $0.23\pm0.11$ \\
\hline
15-30 & 22.5 & $0.53\pm0.09$ & 20-40 & 30 & $0.36\pm0.11$ \\
\hline
30-60 & 45 & $0.55^{+0.10}_{-0.08}$ & 40-70 & 55 & $0.47\pm0.14$ \\
\hline
60-100 & 80 & $0.82^{+0.26}_{-0.24}$ & 70-100 & 85 & $0.61\pm0.15$ \\
\hline
100-150 & 120 & $0.95^{+0.29}_{-0.31}$ & 100-200 & 135 & $0.68\pm0.17$ \\
\hline
--- & --- & --- & 200-300 & 235 & $0.53\pm0.39$ \\
\hline
\hline
CCFR & & & Foudas~et~al. & & \\
$E_\nu$ & $<E_\nu>$ & $\sigma_{l^-l^+}/\sigma_{CC}$ & $E_\nu$ & $<E_\nu>$ & $\sigma_{l^-l^+}/\sigma_{CC}$  \\
(GeV)  & (GeV) & (\%) & (GeV) & (GeV) & (\%) \\
\hline
--- & 40 & $0.54\pm0.05$ & 10-70 & 50 & $0.63\pm0.11$ \\
\hline
--- & 70 & $0.64\pm0.04$ & 70-100 & 80 & $0.60\pm0.07$ \\
\hline
--- & 90 & $0.66\pm0.04$ & 100-150 & 120 & $0.75^{+0.07}_{-0.04}$ \\
\hline
--- & 110 & $0.65\pm0.03$ & 150-200 & 170 & $0.77\pm0.04$ \\
\hline
--- & 140 & $0.77\pm0.04$ & 200-250 & 220 & $0.84\pm0.07$ \\
\hline
--- & 180 & $0.84\pm0.03$ & 250-300 & 270 & $0.86^{+0.09}_{-0.07}$ \\
\hline
--- & 220 & $0.86\pm0.03$ & 300-350 & 320 & $0.99^{+0.09}_{-0.11}$ \\
\hline
--- & 270 & $0.92\pm0.03$ & 350-400 & 370 & $0.97^{+0.15}_{-0.13}$ \\
\hline
--- & 330 & $0.89\pm0.05$ & 400-600 & 450 & $1.10\pm0.18$ \\
\hline
--- & 450 & $0.95\pm0.05$ & --- & --- & --- \\
\hline
\hline

\hline
\hline
\end{tabular}
}
\end{center}
\caption{Di-lepton neutrino cross-section, normalized to the CC cross-section, as a function of the neutrino energy, for some~(\cite{cdhs} to~\cite{ccfr},~\cite{foudas}) electronic detector experiments.}
\label{tab:dile2}
\end{table}

\begin{table}[tbp]
\begin{center}
{\small
\begin{tabular}{||c|c|c|c|c|c||}
\hline
Gargamelle I & & & Gargamelle II & & \\
$E_\nu$ & $<E_\nu>$ & $\sigma_{l^-l^+}/\sigma_{CC}$ & $E_\nu$ & $<E_\nu>$ & $\sigma_{l^-l^+}/\sigma_{CC}$  \\
(GeV)  & (GeV) & (\%) & (GeV) & (GeV) & (\%) \\
\hline
15-35 & --- & $0.65\pm0.23$ & 10-30& 20 & $0.28^{+0.14}_{-0.16}$ \\
\hline
35-75 & --- & $0.69\pm0.22$ & 30-70 & 50 & $0.43\pm0.12$ \\
\hline
75-300& --- & $0.89\pm0.23$ & 70-150 &110 & $0.65^{+0.32}_{-0.30}$ \\
\hline
\hline
E53A+E53B & & & E616+E701 & & \\
$E_\nu$ & $<E_\nu>$ & $\sigma_{l^-l^+}/\sigma_{CC}$ & $E_\nu$ & $<E_\nu>$ & $\sigma_{l^-l^+}/\sigma_{CC}$  \\
(GeV)  & (GeV) & (\%) & (GeV) & (GeV) & (\%) \\
\hline
0-25 & 12.5 & $0.20\pm0.04$ & 30-100 & 70 & $0.48\pm0.06$ \\
\hline
25-50 & 37.5  & $0.51\pm0.08$ & 100-180 & 150 & $0.81\pm0.10$ \\
\hline
50-100& 75& $0.60\pm0.10$ & 180-230 & 200 & $0.90\pm0.10$ \\
\hline
100-200  & 150 & $0.60\pm0.15$ & --- & --- & --- \\
\hline
\hline

\hline
\hline
\end{tabular}
}
\end{center}
\caption{Di-lepton neutrino cross-section, normalized to the CC cross-section, as a function of the neutrino energy,
for some~(\cite{garga1} to~\cite{e53} and~\cite{e616}) electronic detector experiments.}
\label{tab:dile3}
\end{table}

\begin{table}[tbp]
\begin{center}
{\small
\begin{tabular}{||c|c|c|c|c|c||}
\hline
CDHS  & & & CCFR & & \\
$E_\nu$ & $<E_\nu>$ & $\sigma_{l^+l^-}/\sigma_{CC}$ & $E_\nu$ & $<E_\nu>$ & $\sigma_{l^+l^-}/\sigma_{CC}$  \\
(GeV)  & (GeV) & (\%) & (GeV) & (GeV) & (\%) \\
\hline
30-40 & 35 & $0.30\pm0.11$ & --- & 40 & $0.58\pm0.09$ \\
\hline				     
40-60 & 50 & $0.53\pm0.12$ & --- & 70 & $0.72\pm0.07$ \\
\hline				     
60-80 & 70 & $0.71\pm0.15$ & --- & 90 & $0.73\pm0.05$ \\
\hline
80-100 & 90 & $0.78\pm0.17$ & --- & 110&$0.95\pm0.08$ \\
\hline
100-120 & 110 & $0.78\pm0.17$ & ---&140& $0.88\pm0.07$ \\
\hline
120-140 & 130 & $0.83\pm0.18$ & --- & 180 & $1.05\pm0.08$ \\
\hline
140-160 & 150 & $0.86\pm0.19$ & --- & 220 & $1.08\pm0.11$ \\
\hline					 
---     & --- & ---           & --- & 270 & $0.97\pm0.11$ \\
\hline					 
---     & --- & ---           & --- & 330 & $1.30\pm0.22$ \\
\hline					 
---     & --- & ---           & --- & 450 & $0.97\pm0.27$ \\
\hline
\hline
E616+E701 & & & Foudas~et~al. & & \\
$E_\nu$~(GeV) & $<E_\nu>$ & $\sigma_{l^+l^-}/\sigma_{CC}$ & $E_\nu$ & $<E_\nu>$ & $\sigma_{l^+l^-}/\sigma_{CC}$  \\
(GeV)  & (GeV) & (\%) & (GeV) & (GeV) & (\%) \\
\hline
30-100 & 60 & $0.38\pm0.13$ & 12-65 & 50 & $0.63\pm0.15$ \\
\hline
100-230 & 150 & $0.85\pm0.27$ & 65-95 & 80 & $0.84\pm0.15$ \\
\hline
--- & --- &             ---  & 95-145 & 120 & $0.72\pm0.12$ \\
\hline	      
--- & --- &---               & 145-195 & 170 & $0.90\pm0.12$ \\
\hline	       
--- & --- &---               & 195-290 & 230 & $0.84\pm0.12$ \\
\hline	       
--- & --- &---               & 290-600 & 350 & $1.25\pm0.30$ \\
\hline
\hline

\hline
\hline
\end{tabular}
}
\end{center}
\caption{Di-lepton anti-neutrino cross-section, normalized to the CC cross-section, as a function of the neutrino energy, for the electronic detector experiments.}
\label{tab:dile4}
\end{table}

%
%%%%%%%%%%%%%%%%%%%%%%%%%%%%%%%%%%%%%%%%%%%%%%%%%%%%%%%%%%%%%%%%%%%%%%%%%%%%%%
% The world average dilepton cross-section
%%%%%%%%%%%%%%%%%%%%%%%%%%%%%%%%%%%%%%%%%%%%%%%%%%%%%%%%%%%%%%%%%%%%%%%%%%%%%%%
\subsection{The average di-lepton charm-production cross-section}

In order to combine all di-lepton data (see
Tables~\ref{tab:dile2},~\ref{tab:dile3} and~\ref{tab:dile4}) it is
necessary to make a bin by bin weighted average.  Since different
experiments have different binning, we have to define weighting
criteria to combine them all into an unique re-binning of the neutrino
energy range: 0-600 GeV divided into 60 bins. The data-point in the
$j-\mbox{th}$ bin ($(X_j,Y_j), j\in\{1,\ldots,60\}$) is calculated once
we have assigned a given weight to the data-point $(x_{ik},y_{ik})$ of
the $k-\mbox{th}$ bin of the $i-\mbox{th}$ experiments, here-after called source bin, according to the following criteria:

\begin{itemize}
\item the wider the source-bin ($ \Delta x_{ik} $ ), the smaller the weight;
\item the wider the error ($ \Delta y_{ik}$ ), the smaller the weight;
\item the larger the fraction $f_{ijk}$ of the $j-\mbox{th}$ bin
covered by the source bin, the larger the weight.
\end{itemize}

It is then possible to make a bin by bin weighted average of the
data-points by using the following formulae:

\begin{itemize}

\item data-point weight
\begin{displaymath}
  {%
    \omega_{ijk} = \Big(\frac{f_{ijk}}{\Delta y_{ik} \cdot \Delta x_{ik}}\Big)^{2} 
    }
\end{displaymath}

\item data-point average 
\begin{displaymath}
  {%
    \bar{Y}_j = \frac{\sum_{i,k}{y_{ik} \cdot \omega_{ijk}}}{\sum_{i,k}{\omega_{ijk}}} 
    }
\end{displaymath}

\item data-point error
 \begin{displaymath}
   {%
     \Delta \bar{Y}_j = \frac{\sqrt{\sum_{i,k}{\Delta y_{ik}^{2} \cdot \omega_{ijk}^{2}}}}{\sum_{i,k}{\omega_{ijk}}}
     }
\end{displaymath}

\end{itemize}

Since CCFR data do not report the energy bins, it is not
possible to combine them with the others at this stage of the
analysis. 

Once the average has been computed, the CCFR data are combined with
the other ones in the following way: each CCFR data-point is assigned
to the bin containing its x-value, then a standard weighted average
between the CCFR data-point and the previous combined data is
calculated. In this case, the inverse of the $\Delta y$ squared is
used as weight.

\begin{figure}
\begin{center}
\epsfig{file=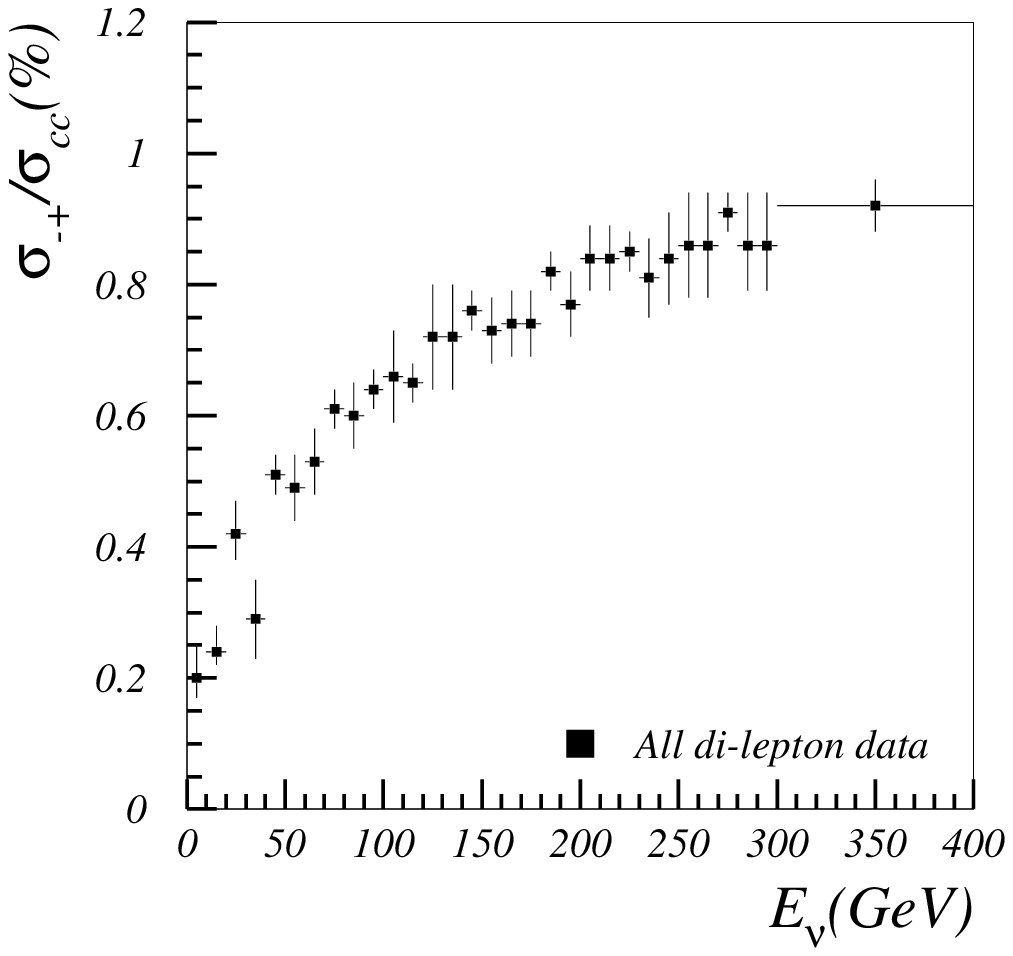,height=12cm,width=12cm}
\end{center}
\caption{\label{fi:diletotcross}Average neutrino di-lepton cross-section, normalized to CC  cross-section, as a function of the neutrino energy}
\end{figure}

\begin{figure}
\begin{center}
\epsfig{file=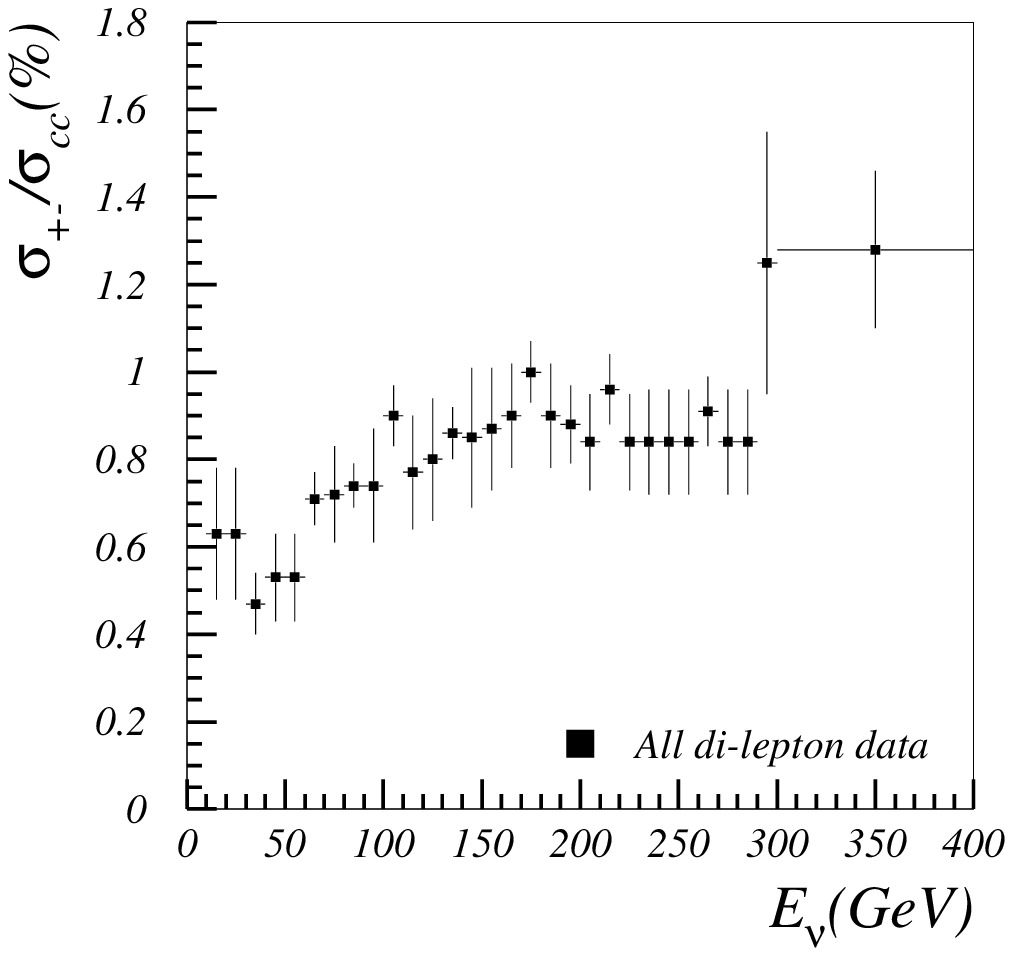,height=12cm,width=12cm}
\end{center}
\caption{\label{fi:antidiletotcross}Average anti-neutrino di-lepton cross-section, normalized to CC  cross-section, as a function of the neutrino energy.}
\end{figure}

The average di-lepton single-charm production cross-sections for both
$\nu$ and $\bar{\nu}$ are shown in
Tables~\ref{tab:diletotcross} and~\ref{tab:antidiletotcross}, and
Figs.~\ref{fi:diletotcross} and~\ref{fi:antidiletotcross}, respectively.

\begin{table}[tbp]
\begin{center}
{\small
\begin{tabular}{||c|c|c|c|c|c||}
\hline
Energy & ${\sigma_{l^-l^+}}/{\sigma_{CC}}$ & Energy  & ${\sigma_{l^-l^+}}/{\sigma_{CC}}$ & Energy  & ${\sigma_{l^-l^+}}/{\sigma_{CC}}$ \\
 (GeV) & (\%) & (GeV) & (\%) & (GeV) & (\%) \\
\hline \hline
   $5$   & $0.20^{+0.05}_{-0.03}$ & $145$ & $0.76\pm0.03$ & $285$ & $0.86\pm0.08$  \\
\hline							                         
   $15$  & $0.24^{+0.04}_{-0.02}$ & $155$ & $0.73\pm0.05$ & $295$ & $0.86\pm0.08$   \\
\hline							                         
   $25$  & $0.42^{+0.05}_{-0.04}$ & $165$ & $0.74\pm0.05$ & $350$ & $0.92\pm0.04$  \\
\hline				                       	  			 
   $35$  & $0.29\pm0.06        $ & $175$ & $0.74\pm0.05$ & $450$ & $0.96\pm0.05$   \\
\hline				                       	                          
   $45$  & $0.51\pm0.03$         & $185$ & $0.82\pm0.03$ & $550$ & $1.10\pm0.18$   \\
\hline			   	                       	                          
   $55$  & $0.49\pm0.05$         & $195$ & $0.77\pm0.05$ &  & \\
\hline			   	                       	   
   $65$  & $0.53\pm0.05$         & $205$ & $0.84\pm0.05$ &  & \\
\hline			   	  			    
   $75$  & $0.61\pm0.03$         & $215$ & $0.84\pm0.05$ &  & \\
\hline			   	                       	    
   $85$  & $0.60\pm0.05$         & $225$ & $0.85\pm0.03$ &  & \\
\hline			   	                       	   
   $95$  & $0.64\pm0.03$         & $235$ & $0.81\pm0.06$ &  & \\
\hline			   	                       	  
   $105$ & $0.66\pm0.07$         & $245$ & $0.84\pm0.07$ & & \\
\hline				                       	  
   $115$ & $0.65\pm0.03$         & $255$ & $0.86\pm0.08$ & &\\
\hline				                       	  
   $125$ & $0.72\pm0.08$         & $265$ & $0.86\pm0.08$ & &\\
\hline				                          
   $135$ & $0.72\pm0.08$         & $275$ & $0.91\pm0.03$ &  &\\
\hline
\hline
\end{tabular}}
\end{center}
\caption{Average neutrino di-lepton cross-section, normalized to CC cross-section, as a function of the neutrino energy.}
\label{tab:diletotcross}
\end{table}

\begin{table}[tbp]
\begin{center}
{\small
\begin{tabular}{||c|c|c|c|c|c||}
\hline
Energy & ${\sigma_{l^+l^-}}/{\sigma_{CC}}$ & Energy & ${\sigma_{l^+l^-}}/{\sigma_{CC}}$ & Energy & ${\sigma_{l^+l^-}}/{\sigma_{CC}}$ \\
 (GeV) & (\%) & (GeV) & (\%) & (GeV) & (\%) \\
\hline \hline
   $5$   & --- & $145$ & $0.85\pm0.16$ & $285$ & $0.84\pm0.12$  \\
\hline							                         
   $15$  & $0.63\pm0.15$ & $155$ & $0.87\pm0.14$ & $295$ & $1.25\pm0.30$   \\
\hline							                         
   $25$  & $0.63\pm0.15$ & $165$ & $0.90\pm0.12$ & $350$ & $1.28\pm0.18$  \\
\hline				                       	  			 
   $35$  & $0.47\pm0.07$ & $175$ & $1.00\pm0.07$ & $450$ & $1.10\pm0.20$   \\
\hline				                       	                          
   $45$  & $0.53\pm0.10$ & $185$ & $0.90\pm0.12$ & $550$ & $1.25\pm0.30$   \\
\hline			 	                       	                          
   $55$  & $0.53\pm0.10$ & $195$ & $0.88\pm0.09$ &  & \\
\hline			 	                       	  			 
   $65$  & $0.71\pm0.06$ & $205$ & $0.84\pm0.11$ &  & \\
\hline			 	  		            
   $75$  & $0.72\pm0.11$ & $215$ & $0.84\pm0.11$ &  & \\
\hline			 	                            
   $85$  & $0.74\pm0.05$ & $225$ & $0.96\pm0.08$ &  & \\
\hline			 	                  	 
   $95$  & $0.74\pm0.13$ & $235$ & $0.84\pm0.11$ & &  \\
\hline			 	                          
   $105$ & $0.90\pm0.07$ & $245$ & $0.84\pm0.12$ &  &\\
\hline			 	                          
   $115$ & $0.77\pm0.13$ & $255$ & $0.84\pm0.12$ & &\\
\hline			 	                  
   $125$ & $0.80\pm0.14$ & $265$ & $0.91\pm0.08$ & &\\
\hline			 	                  
   $135$ & $0.86\pm0.06$ & $275$ & $0.84\pm0.12$ &  &\\
\hline
\hline
\end{tabular}}
\end{center}
\caption{Average anti-neutrino di-lepton cross-section, normalized to CC  cross-section, as a function of the neutrino energy.}
\label{tab:antidiletotcross}
\end{table}

%
%%%%%%%%%%%%%%%%%%%%%%%%%%%%%%%%%%%%%%%%%%%%%%%%%%%%%%%%%%%%%%%%%%%%%%%%%%%%%%
% Emulsion experiments
%%%%%%%%%%%%%%%%%%%%%%%%%%%%%%%%%%%%%%%%%%%%%%%%%%%%%%%%%%%%%%%%%%%%%%%%%%%%%%%
\section{Charm-production studies with nuclear emulsions}
\label{emu}
\subsection{Experimental issues}
So far only two experiments, E531~\cite{e531} and CHORUS~\cite{chorus},
have searched for inclusive charm-production through the direct
identification of charm decays in the emulsions. The main advantage of
these experiments is that, being the charmed particle identified
through its decay, very loose kinematical cuts are applied. This
translates into a very good sensitivity to the slow-rescaling threshold
behavior and consequently to the charm-quark mass.

These experiments have a hybrid design. Electronic detectors locate
the regions in emulsions where the neutrino interactions occur and contribute to the
reconstruction of the event kinematics. Emulsions are used as active
targets. They have the appropriate position resolution (less than
1~$\mu$m) and granularity to detect short-lived particles through the
visual observation of their decays. The main background for $D^0$
detection in emulsion comes from $K^0$ and $\Lambda$ decays, and
neutron, $K^0$ and $\Lambda$ interactions without any visible nuclear
break-up at the interaction point. For charged charmed-hadrons the main
backgrounds are $\pi$ and $K$ decays in flight, and the {\em white
kink} (hadron interaction without any visible nuclear break-up) on any
charged non-charmed hadron. Both the processes can be easily ruled out
by applying a cut on the transverse momentum at the decay vertex
($p_T>250~\mbox{MeV}$). The contribution to the background of these
processes is of the order of $10^{-4}/CC$.

\subsection{Emulsion data sample and selection criteria}

Although in the last years the automatic scanning of the nuclear
emulsions had had an impressive development, the event search principle
used in the former hybrid experiment E531 is still valid for present
experiments. Once the neutrino interaction is successfully
reconstructed, tracks are searched for in the downstream end of
the emulsion target by using the slopes and coordinates as predicted
by the electronic detectors. If the predicted track is found in the
most downstream emulsion sheet, it is then followed through the
emulsion module up to the neutrino interaction vertex.

Once a neutrino interaction is found, an attempt is made to find
possible charmed-particle decays. In the former E531 experiment
semi-automatic systems were used, thus limiting the number of
interactions to be studied. Nowadays, very powerful automatic systems
are available allowing for a larger statistics to be searched for.

The final E531 data sample consists of 3855 neutrino interactions among
which 122 charmed-particle decays are found~\cite{e531}. The
estimated background is 0.2 and 3.6 events for neutral and charged
charmed-particle decays, respectively.

Using the weighted number of events, the relative
charmed-particle production rate is: $\sigma(\nu_\mu N\rightarrow
c\mu^-X)/\sigma(\nu_\mu N\rightarrow \mu^-X)=4.9^{+0.7}_{-0.6}\%$.
Table~\ref{tab:emucross} shows the energy dependence of this
cross-section. The analysis of the CHORUS data is still under way.

\begin{table}[tbp]
\begin{center}
{\small
\begin{tabular}{||c|c|c||}
\hline
$E_\nu$ & $<E_\nu>$ & $\sigma_C/\sigma_{CC}$ \\
(GeV) & (GeV) & (\%)\\
\hline
0-10 & 5 & $0.7^{+1.1}_{-0.6}$ \\
\hline
10-30 & 20 & $3.6\pm0.9$ \\
\hline
30-60& 45 & $4.9\pm1.1$ \\
\hline
60-150& 105 & $10.1\pm2.5$ \\

\hline
\hline
\end{tabular}
}
\end{center}
\caption{Inclusive neutrino charm-production cross-section, normalized to the CC cross-section, as a function of the neutrino energy.}
\label{tab:emucross}
\end{table}

%In the following we describe the Netscan analysis used in CHORUS to
%search for decay topologies. This technique, originally developed for
%the DONUT experiment, is based on the reconstruction of all tracks in
%a given fiducial volume around the neutrino interaction vertex.

%Once the interaction vertex is found, according to the procedure
%described in~\cite{location}, a Netscan is performed in a volume of
%$1.5\times1.5\times\times6.28~\mbox{mm}^3$ centered on the assumed
%vertex position. All tracks segments in the fiducial volume with an
%angle less than 400~mrad are recorded and used as an input for an
%off-line program which performs tracks and vertices reconstruction.

%
%%%%%%%%%%%%%%%%%%%%%%%%%%%%%%%%%%%%%%%%%%%%%%%%%%%%%%%%%%%%%%%%%%%%%%%%%%%%%%
% Combining di-lepton and emulsion data 
%%%%%%%%%%%%%%%%%%%%%%%%%%%%%%%%%%%%%%%%%%%%%%%%%%%%%%%%%%%%%%%%%%%%%%%%%%%%%%%
\section{The world average charm-production cross-section}
\label{combi}

\subsection{From the di-lepton to the total charm-production cross-section}
As discussed in Section~\ref{dilepto}, calorimetric and bubble chamber
experiments cannot detect directly the charmed hadrons; instead only
the lepton from semi-leptonic charm decay is measured. Without going
into the details, we remind that the charmed hadron (C) cross-section
can be connected to the charmed quark (c) cross-section via
fragmentation functions, assuming factorization~\cite{bolton}:

\begin{eqnarray}
\frac{d\sigma(\nu N\rightarrow\mu^- C X)}{dxdydzdp_T^2} & = &\frac{d\sigma(\nu N\rightarrow\mu^- c X)}{dxdy}\times\sum_h{f_h\times D_c^h(z,p_T^2)} \nonumber \\ 
\frac{d\sigma(\bar{\nu} N\rightarrow\mu^+ \bar{C} X)}{dxdydzdp_T^2} & = & \frac{d\sigma(\nu N\rightarrow\mu^+ \bar{c} X)}{dxdy}\times\sum_h{\bar{f}_h\times \bar{D}_{\bar{c}}^h(z,p_T^2)} 
\end{eqnarray}

Here, $D_c^h(z,p_T^2)$ is the probability distribution for the
charmed-quark to fragment into a charmed hadron of type $h (= D^0,
D^+, D_s^+, \Lambda_c^+)$ carrying a fraction $z$ of the quark
longitudinal momentum and transverse momentum $p_T$ with respect
to the quark direction. The number $f_h$ is the mean multiplicity of
the hadron $h$ in neutrino charm-production. Analogously for
anti-neutrinos, indicated by the barred quantities. Since only one
$c$-quarks is produced in a CC interaction, one can set the
normalization conditions as

\begin{equation}
\int_0^1dz\int_0^\infty  D_c^h(z,p_T^2)dp_T^2 = \int_0^1dz\int_0^\infty \bar{D}_{\bar{c}}^{\bar{h}}(z,p_T^2)dp_T^2 = 1
\end{equation}

and

$$ \sum_h f_h = \sum_h\bar{f}_h = 1\,.$$

The total cross-section for di-lepton charm-production, at a given
energy, can then be written as

\begin{eqnarray}
\sigma(\nu N\rightarrow\mu^-l^+ X)(E) & = &\sigma(\nu N\rightarrow\mu^-
C X)(E)\times \sum_h {f_h(E)\times BR_l^h} \nonumber \\
 & = & \sigma(\nu N\rightarrow\mu^- C X)(E)\times\overline{BR}_l(E)
\end{eqnarray}

where $BR_l^h$ is the semi-leptonic branching fraction for the charmed
hadron $h$.

\begin{table}[tbp]
\begin{center}
{\small
\begin{tabular}{||c|c|c|c|c||}
\hline
Energy (GeV) & $f_{D^0}$ & $f_{D^+}$ & $f_{D^+_s}$ & $f_{\Lambda^+_c}$ \\
\hline \hline
$5-20$ & $0.32\pm0.11$ & $0.05\pm0.06$ & $0.18\pm0.10$ & $0.44\pm0.12$ \\
 \hline
$20-40$ & $0.50\pm0.08$ & $0.10\pm0.08$ & $0.22\pm0.08$ & $0.18\pm0.07$ \\
 \hline
$40-80$ & $0.64\pm0.08$ & $0.22\pm0.09$ & $0.09\pm0.08$ & $0.05\pm0.04$ \\
 \hline
$>80$ & $0.60\pm0.11$ & $0.30\pm0.11$ & $0.00\pm0.06$ & $0.09\pm0.08$ \\
\hline
\hline
\end{tabular}}
\end{center}
\caption{E531 prediction fraction results as obtained in Ref.~\cite{bolton}.}
\label{tab:fhenedep}
\end{table}

From the previous formulae it is clear that in order to go from the
di-lepton charm-production to the inclusive charm-production, $BR_l^h$
and $f_h(E)$ should be known. The semi-leptonic branching fractions
are measured with an accuracy of about 10\%~\cite{pdg}, while the
$f_h(E)$'s have been measured directly only in one experiment,
E531. However, a bias was found in the way E531 extracted $f_h(E)$. A
detailed discussion on the re-fitting of these data with the bias
removed is reported in Ref.~\cite{bolton}. For our purposes the measured
energy dependence of $f_h$ is relevant, see Table~\ref{tab:fhenedep}.
The error on $\overline{BR}_l(E)$ has been computed as :

$$
(\Delta \overline{BR}_l(E))^2 = \sum_i f_i^2(E) \cdot (\Delta BR_i)^2 + \sum_i BR_i^2 \cdot (\Delta f_i(E))^2 + 2 \cdot \sum_{j>i} BR_i \cdot BR_j \cdot cov_{j,i}(E)
$$

where $cov_{j,i}(E)$ is the covariance matrix of the $f_h$ values,
defined as $${ cov_{i,j}(E)=\Delta f_i(E) \cdot \Delta f_j(E) \cdot
cor_{i,j}(E) }$$ and $cor_{i,j}(E)$ is the correlation among the
$f_h$~\cite{bolton}.

The average inclusive neutrino charm-production cross-section,
normalized to the CC cross-section, as a function of the neutrino
energy is shown in Table~\ref{tab:dileinclu} and
Fig.~\ref{alldilefin}.

\begin{table}[tbp]
\begin{center}
{\small
\begin{tabular}{||c|c|c|c|c|c||}
\hline
Energy & ${\sigma_C}/{\sigma_{CC}}$ & Energy & ${\sigma_C}/{\sigma_{CC}}$ & Energy & ${\sigma_C}/{\sigma_{CC}}$ \\
(GeV) & (\%) &  (GeV) & (\%) & (GeV) & (\%) \\
\hline \hline
   $5$   & $3.12^{+1.01}_{-0.71}$ & $145$ & $8.00\pm1.63$ & $285$ & $9.05\pm2.00$  \\
\hline							                         
   $15$  & $3.75^{+0.99}_{-0.72}$ & $155$ & $7.68\pm1.62$ & $295$ & $9.05\pm2.00$   \\
\hline							                         
   $25$  & $5.53^{+1.46}_{-1.34}$ & $165$ & $7.79\pm1.64$ & $350$ & $9.68\pm1.98$  \\
\hline				                       	  			 
   $35$  & $3.82^{+1.20}_{-1.16}$ & $175$ & $7.79\pm1.64$ & $450$ & $10.11\pm2.09$   \\
\hline				                       	                          
   $45$  & $5.67\pm1.06$         & $185$ & $8.63\pm1.75$ & $550$ & $11.58\pm2.99$   \\
\hline			   	                       	                          
   $55$  & $5.44\pm1.12$         & $195$ & $8.11\pm1.70$ &  & \\
\hline			   	                       	  
   $65$  & $5.89\pm1.19$         & $205$ & $8.84\pm1.85$ &  & \\
\hline			   	  			   
   $75$  & $6.78\pm1.25$         & $215$ & $8.84\pm1.85$ &  & \\
\hline			   	                       	   
   $85$  & $6.32\pm1.37$         & $225$ & $8.95\pm1.82$ &  & \\
\hline			   	                       	  
   $95$  & $6.74\pm1.38$         & $235$ & $8.53\pm1.82$ &  & \\
\hline			   	                       	  
   $105$ & $6.95\pm1.57$         & $245$ & $8.84\pm1.92$ &  &\\
\hline				                       	  
   $115$ & $6.84\pm1.40$         & $255$ & $9.05\pm2.00$ & &\\
\hline				                       
   $125$ & $7.58\pm1.73$         & $265$ & $9.05\pm2.00$ & &\\
\hline				                          
   $135$ & $7.58\pm1.73$         & $275$ & $9.58\pm1.94$ & & \\
\hline
\hline
\end{tabular}}
\end{center}
\caption{Average inclusive neutrino charm-production cross-section,
normalized to the CC cross-section, as a function of the neutrino
energy.}
\label{tab:dileinclu}
\end{table}

%
%%%%%%%%%%%%%%%%%%%%%%%%%%%%%%%%%%%%%%%%%%%%%%%%%%%%%%%%%%%%%%%%%%%%%%%%%%%%%%
\subsection{The world average total charm-production cross-section}
%%%%%%%%%%%%%%%%%%%%%%%%%%%%%%%%%%%%%%%%%%%%%%%%%%%%%%%%%%%%%%%%%%%%%%%%%%%%%%

\begin{figure}
\begin{center}
\epsfig{file=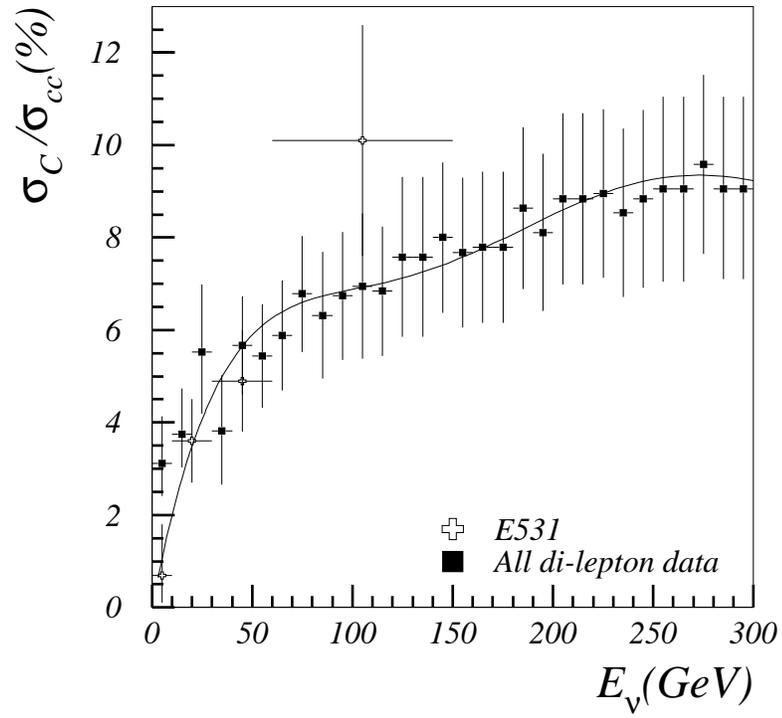,height=12cm,width=12cm}
\end{center}
\caption{\label{alldilefin} Average inclusive neutrino
charm-production cross-section, normalized to the CC cross-section, as
a function of the neutrino energy. The continuous line shows the result
of a fit to the data with a polynomial function.}
\end{figure}

The inclusive charm-production cross-section as measured in the E531
experiment and as derived from di-lepton data are shown in
Fig.~\ref{alldilefin}. The agreement between the two
determinations is quite good, although the statistical errors are
large. A better determination of the charmed fractions in the CHORUS
experiment would reduce significantly the errors on the cross-section
as derived from di-lepton data.

In order to parametrise the inclusive charm-production cross-section
as a function of the energy, we have performed a fit with a polynomial
function ($f(x)$) requiring that the constant term is zero:

$$f(x) = a_1x+a_2x^2+a_3x^3+\ldots+a_nx^n$$

Starting from $n=6$, the $\chi^2$ of the fit is very good, $P(\chi^2)>0.99$, 
and mildly dependent on $n$. Therefore, we have chosen the smallest power for 
$n$. We have studied the effect of the different parametrisations
on the prediction of the inclusive charm-production cross-section and
we have found results different of at most 5\% from the ones obtained
with $n=6$. Therefore,  we have assigned to each prediction a
systematical error of 5\%. The fitted parameters $a_i$ together with
the correlation coefficients $cor(a_i,a_j)$ are given in
Table~\ref{tab:fitresults}. The fitted function is also shown in
Fig.~\ref{alldilefin}. Notice that the present parametrisation holds in the 
range $[0 \div 300]$~GeV.

\begin{table}[tbp]
\begin{center}
{\small
\begin{tabular}{||c||c|cccccc||}
\hline
$a_i\pm\Delta a_i$ & & $a_1$ & $a_2$ & $a_3$ & $a_4$ & $a_5$ & $a_6$  \\
\hline \hline
$0.23\pm0.03$                    & $a_1$ &  1.000 & -0.828 & 0.266 & 0.228 &  0.028 & -0.204 \\ 
$-(0.31\pm0.03)\times 10^{-2}$   & $a_2$ & -0.828 &  1.000 &-0.664 &-0.132 &  0.096 &  0.196\\
$(0.21\pm0.07)\times 10^{-4}$  & $a_3$ &    0.266 & -0.664 & 1.000 &-0.463 & -0.168 &  0.151\\
$-(0.71\pm0.02)\times 10^{-7}$ & $a_4$ &    0.228 & -0.132 &-0.463 & 1.000 & -0.371 & -0.225\\
$(0.115\pm0.003)\times 10^{-9}$   & $a_5$ &    0.028 &  0.096 &-0.168 &-0.371 &  1.000 & -0.735\\
$-(0.71\pm0.05)\times 10^{-13}$   & $a_6$ &   -0.204 &  0.196 & 0.151 &-0.225 & -0.735 &  1.000 \\
\hline
\hline
\end{tabular}}
\end{center}
\caption{Fitted parameters $a_i$ together with the correlation
coefficients $cor(a_i,a_j)$.}
\label{tab:fitresults}
\end{table}

As an application of our fit to the available data, we compute the
expected $D^0$ production, normalised to the neutrino CC
cross-section, in the CHORUS experiment, whose neutrino beam has an
average energy of 27~GeV. By convoluting the neutrino spectrum with
the cross-section shown in Fig.~\ref{alldilefin}, the expected average
inclusive charm-production, normalized to the neutrino CC
cross-section, is $(4.81\pm0.16\mid_{stat}\pm0.24\mid_{syst})\%$,
while the expected $\sigma(D^0)/\sigma_{CC}$ is
$(2.58\pm0.26\mid_{stat}\pm0.13\mid_{syst})\%$. This has to be
compared with the measured value of
$2.34\pm0.15\pm0.17\%$~\cite{guler}. Notice that the former experiment
E531, with an average neutrino energy of about 22~GeV, measured the ratio
$\sigma(D^0)/\sigma_{CC}=2.19^{+0.39}_{-0.35}$~\cite{fredi}.

%
%%%%%%%%%%%%%%%%%%%%%%%%%%%%%%%%%%%%%%%%%%%%%%%%%%%%%%%%%%%%%%%%%%%%%%%%%%%%%%
% Conclusion
%%%%%%%%%%%%%%%%%%%%%%%%%%%%%%%%%%%%%%%%%%%%%%%%%%%%%%%%%%%%%%%%%%%%%%%%%%%%%%%
\section{Conclusion}
Over the past 30 years, several experiments have accumulated data on
the (anti-)neutrino induced single-charm production with different
techniques: opposite-sign dimuons have been used by calorimetric
experiments, di-leptons ($\mu,~e$) by bubble chambers while only
nuclear emulsion experiments have identified the charm through the
visual observation of its decay. By combining all these data, a wide
energy range (0-600~GeV) has been spanned, providing a good
sensitivity to the threshold behavior of the cross-section. \\ We have
presented a statistical way of combining all the available data and
hence the extraction of the world averaged single-charm production
cross-section as a function of the neutrino energy. A comparison of
the $D^0$ production cross-section, normalised to the neutrino CC
cross-section, with the result recently presented by the CHORUS
Collaboration shows a good agreement.

\section*{Acknowledgment}
We gratefully acknowledge U. Dore for useful discussions.
%%%%%%%%%%%%%%%%%%%%%%%%%%%%%%%%%%%%%%%%%%%%%%%%%%%%%%%%%%%%%%%%%%%%%%%%%%%%% 
% Bibliography
%%%%%%%%%%%%%%%%%%%%%%%%%%%%%%%%%%%%%%%%%%%%%%%%%%%%%%%%%%%%%%%%%%%%%%%%%%%%%%
%

%%%%%%%%%%%%%%%%%%%%%%%%%%%%%%%%%%%%%%%%%%%%%%%%%%%%%%%%%%%%%%%%%%%%%%%%%%%%%%

\end{document}